# The Effect of IoT New Features on Security and Privacy: New Threats, Existing Solutions, and Challenges Yet to Be Solved

Wei Zhou, Yuqing Zhang, and Peng Liu, *Member, IEEE*

*Abstract*—The future of Internet of Things (IoT) is already upon us. IoT applications have been widely used in many field of social production and social living such as healthcare, energy and industrial automation. While enjoying the convenience and efficiency that IoT brings to us, new threats from IoT also have emerged. There are increasing research works to ease these threats, but many problems remain open. To better understand the essential reasons of new threats and the challenges in current research, this survey first proposes the concept of "IoT features". Then, the security and privacy effects of eight IoT new features were discussed including the threats they cause, existing solutions and challenges yet to be solved. To help researchers follow the up-to-date works in this field, this paper finally illustrates the developing trend of IoT security research and reveals how IoT features affect existing security research by investigating most existing research works related to IoT security from 2013 to 2017.

## I. Introduction

With the development of critical technologies in the Internet of things (IoT), the IoT applications (e.g., smart home, digital healthcare, smart grid, smart city) become widely used in the world. According to statistics website Statista [1], the number of connected devices around the world will dramatically increase from 20.35 billion in 2017 to 75.44 billion in 2025. International Data Corporation (IDC) [2] has predicted a 17.0% compound annual growth rate (CAGR) in IoT spending from $698.6 billion in 2015 to nearly $1.3 trillion in 2019, there seems to be a consensus that the impact of IoT technologies is substantial and growing.

Along with the rapid growth of IoT application and devices, cyber-attacks will also be improved and pose a more serious threat to security and privacy than ever before. For instance, remote adversaries could compromise patients' Implantable medical devices [3] or smart cars [4], which may not only cause huge economic losses to individuals but also threat peoples' lives. Furthermore, as the IoT devices become widely used in industry, military, and other key areas, hackers can jeopardize public and national security. For example, on 21 October 2016, a multiple distributed denial of service (DDoS) [5] attacks systems operated by Domain Name System provider Dyn, which caused the inaccessibility of several websites such as GitHub, Twitter, and others. This attack just executed through a botnet consisting of a large number of IoT devices including printers, IP cameras, gateways and baby monitors etc. In another instance, Stuxnet [6] a malicious computer worm targets industrial computer systems were responsible for causing substantial damage to Iran's nuclear program.

However, most of the enterprises and individuals lack awareness of privacy and security. A recent study by Pew Research Center [7] found that many Americans feel over-optimistic about how their data have been used. Only 26% Americans do not accept their health information to be shared with their doctor. Moreover, nearly half of Americans agreed that it was acceptable auto insurance companies to monitor their location and driving speed in order to offer discounts on their insurance. On the other hand, due the lack of customer demand, manufacturers used to focus on implementing products' core functions while ignoring security. Meanwhile, IoT devices vendors generally do not send updates and patches to their devices unless user-initiated firmware updates. At the same time, IoT devices typically do not run full-fledged security mechanisms due to constrained consumption and resource. As a result, IoT devices often remain easy-to-use vulnerabilities (e.g., default passwords, unpatched bugs) for extended periods [8].

Motivated by an increasing number of vulnerabilities, attacks and information leaks, IoT device manufactures, cloud providers, and researchers are working to design systems to security control the flow of information between devices, to detect new vulnerabilities, and to provide security and privacy within the context of users and the devices. While researchers continue to tackle IoT security and privacy, the most studies are only in its incipient stages and lack applicability, and many problems remain open. In order to point out valuable directions for further research and provide useful references for researchers, there are many published survey on IoT security. Li *et al.* [9] and Lin *et al.* [10] mainly discussed and analyzed current attacks and challenges following layers. Fu *et al.* [11] highlight some opportunities and potential threats in two different application scenarios-home and hospital. Roman *et al.*

W. Zhou, Y. Zhang are with the National Computer Network Intrusion Protection Center, University of Chinese Academy of Sciences, Beijing 100000, China (e-mail: zhouw@nipc.org.cn; zhangyq@nipc.org.cn).

P. Liu is with the College of Information Sciences and Technology, The Pennsylvania State University, PA 16802, USA, (e-mail: pliu@ist.psu.edu).

[12] and Sicari *et al.* [13] presented research challenges and the promising solutions focusing on different features and security mechanism including authentication, access control, confidentiality, privacy. The latest survey published by Yang *et al.* [14] synthesis main point of previous surveys and present the classification of IoT attacks. They all presented most aspects of IoT security research, threats, and open issues, and suggest some hints for future research. However, few of them exposed and deeply analyzed the root cause of these challenges and threats, and clearly identify what new challenges coming from IoT. Although Yang *et al.* and Trappe *et al.* [15] discussed some relevant limitations of IoT devices, they just focus on the challenges caused by restricted battery capacity and computing power. There are many more IoT constraints and features have not been covered could affect the security and privacy.

To fill the gap, this paper discusses and analyzes the IoT security issues from a new perspective - IoT features. "IoT features" refers to the unique features of IoT devices network and applications, which are different with traditional Internet and computers. For example, IoT devices have much less computing ability, storage resources, and power supply, thus "Constrained" is seen as an IoT feature. The contribution of this paper can be summarized as follows:

- *a).* To find out the basic cause of current IoT threats and main challenges in IoT research, we first time propose the concept of "IoT features".
- *b).* To better understand the effect of IoT features, we describe eight features which have most impact on security and privacy issues and discuss the threats, research challenges, and opportunities derived from each feature.
- *c).* We present the development trends of current IoT security and its cause based on IoT features though the analysis of existing research in recent five years.

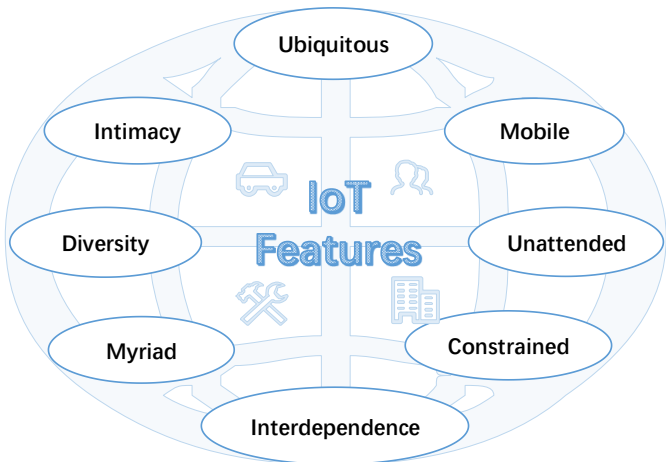

Fig. 1. IoT Features.

The rest of paper is organized as follows. Section II is the main parts of this paper, we focus on eight IoT features as shown in Fig. 1, and fully discuss and analyze them respectively. Then we collect nearly 200 research related to IoT security from 2013 to 2017 and provide many kinds of statistical analysis with them in Section III. Finally, conclusions are presented in Section IV.

## II. THE EFFECT OF IOT FEATURES ON SECURITY AND PRIVACY

In this section, we will elaborate four aspects about each IoT features in Fig 1: description, threat, challenges, solutions, and opportunities.

1). *Description:* We introduce what the feature is and what the differences between traditional devices, network, and applications are.
2). *Threat:* We discuss what potential threats and vulnerabilities brought by the feature, and the consequences caused by these threats. We also provide diagrams and attack examples for some threats, which makes it easy to follow.
3). *Challenges:* We present what research challenges caused by the features.
4). *Solutions & Opportunities:* We present existing solutions to tackle the challenges and the drawbacks of these solutions. In addition, we also introduce some new security techniques/ideas that could also help to migrate the challenges and threats as opportunities here.

### A. Interdependence

*1) Description:* As the number of IoT devices increases, the interaction between devices become more complex and need less human involvement. IoT devices are no longer just communicate explicitly with each other like traditional computers or smartphones. Many of them could also implicitly controlled by other multiple devices behaviors or environmental conditions using services like IFTTT (if this then that) [16], which is popular in various IoT application scenarios. For example, if the thermometer detects the indoor temperature has been raised and the threshold and smart plug detect the air conditioner was in the "off" state, and then the windows would automatically open. The similar examples are more common in industrial and agricultural devices (e.g., automatic adding more water into smelters according to temperature and humidity). We call this implicitly dependence relationship between devices as an IoT feature named "Interdependence" here.

*2) Threats:* The target device or system itself might not be easily compromised, but the attackers could easily change other devices behavior or the surrounding environment, which have interdependence relationship to achieve their aims. As a result, this feature could be maliciously used to reduce the difficulty of direct attack the target devices and bypass original defense mechanism. For example, back to the scenario described as the first example in the last paragraph, the hackers do not need to attack the automatic window control or thermometer. However, he could compromise the smart plug that connected to the public network to turn off the air-conditioner in a room and trigger a temperature increase, which would result in the windows to open and create a physical security breach, as shown in Fig. 2.

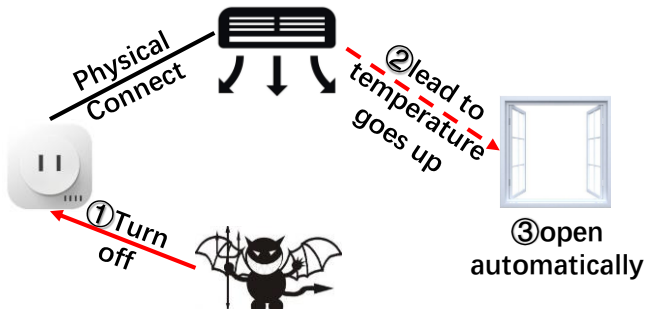

Fig. 2. Attack Example of Interdependence behaviors.

*3) Challenges:* The majority of the researchers do not realize the effect of interdependence behaviors on IoT security. Researchers generally protect the single device itself. However, it is difficult to make a clear defensive boundary of IoT devices or use static access control methods and privilege management to them because of their interdependent behaviors. In addition, the management of most of IoT devices controlled by cloud platforms applications (e.g., Samsung SmartThings [17], Apple HomeKit [18], Amazon Alexa [19], JD [23], and Ali [24]), which have already gained great popularity among smart home users today. Due the IoT device behaviors could be changed with other devices or environmental conditions, it is difficult to define a certain set of fine-grained permission rules for them. The *overprivilege* has become a common problem in the permission model of existing IoT platforms applications [20].

*4) Solutions & Opportunities:* The team at Carnegie Mellon University was aware of the cross-device dependencies early, and proposed a set of new security policies for detecting anomaly behavior of interdependence [21]. However, these policies will be more complicated and impractical with the increasing number of devices. Last year, Yunhan *et al.* [22] proposed *ContexIoT*, a new context-based permission system for IoT platforms application to solve overprivileged problem. It records and compares more context information such as procedure control and data flow, and runtime data of every IoT device action before it is executed, and then let the user allow or deny this action based on this information. This method could detect the misuse of IoT devices interdependence behaviors as early. Because even if hackers make the misbehavior at the same physical conditions with the normal, it is hard to create the same context information like data sources. However, this method still too dependent on user decision, so once user makes a wrong decision, the system will remember this wrong decision and will not prompts the user again. While more effective and practical solutions are urgently needed to address the threats posed by the interdependence.

### B. Diversity

*1) Description:* On the one hand, as IoT technology widely used in more application scenarios. More kinds of IoT devices are designed for specific tasks and interact strongly with the physical environment. Thus, their hardware, system, and process requirements are unique. For example, a small temperature sensor might run on a single chip MCS-51 with a few KB flash and RAM, while a complex machine tool might have higher performance than our smartphone. On the other hand, in different application scenarios also need different network and communication protocols. To seize the IoT market, many large IT companies launched their cloud platform to manage IoT devices as we mentioned above, and each of them designs their own wireless access, authentication and communication protocols. We call the many different kinds of IoT devices and protocols as an IoT feature named "diversity" here

*2) Threats:* Due to increasing kinds of new IoT devices began flooding the IoT market with fewer safety checks, Ali mobile security team [25] found more than 90% of IoT device firmware has security vulnerabilities like hard-coded key, and 94% known Web security vulnerabilities still existed in these devices' Web interfaces, which could easily be used by hackers.

In order to roll out IoT cloud platform quickly and lacking the experience for new IoT application demand such as IoT device bootstrapping [26], the protocols designed by IT companies may have many potential security problems. For instance, Liu *et al.* [27] found the attack could carry out several attacks with JoyLink protocol of JD, such as device hijacking shown in Fig. 3. Moreover, different protocols have different semantic definitions, the attackers also could use this point to find security vulnerabilities like *BadTunnel* [28] when they uncorrected work together.

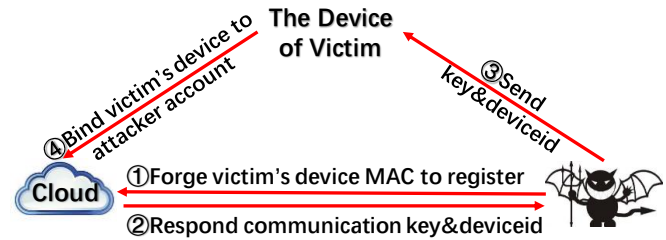

Fig. 3. Device Hijacking Attack Example of JoyLink Protocols.

*3) Challenges:* For system security, due to the diversity of IoT devices, it is hard to design a common system defense for the heterogeneous devices, especially in industry area [29]. Thus, how to discover and deal with so many security vulnerabilities among the various IoT devices needs to be addressed urgently.

For network security, due every protocol has differences with others, so it is important for researchers to dig out general crucial security problems of them. Besides, the security problems for the protocol and network themselves, researchers should also consider the potential security issues caused by association with different protocols.

*4) Solutions & Opportunities:* To discover and address the potential vulnerabilities for more kinds of IoT devices, researchers attempted to use static or dynamic analysis [30] of the firmware and source running on these devices. In 2014, Zaddach *et al.* [31] put forward a framework to support dynamic security analysis for a variety of embedded systems' firmware. It cannot simulate all action of the real devices and need to forward action from the emulator to the device. Thus, it is unsuitable for large-scale firmware analysis without physical connecting devices. Chen *et al.* [32] presented a framework for large-scale automated firmware dynamic analysis, but it is only applicable to the Linux-based system. The full firmware dynamic analysis simulation framework for Real-Time Operating System (RTOS) and bare-metal system is nearly blank.

On the other hand, researchers rely on the Intrusion Detection System (IDS) and intrusion prevention system (IPS) to protect many kinds of devices at same time. However, the different attacks vary according to their target devices, thus some researchers pointed out the IDS and IPS systems model based on anomaly traffic detection may not work well to the different kinds of devices. They suggested that the IDS and IPS systems should first detect abnormal the parameter which could affect the devices' behaviors among network traffic. For example, Hadziosmanovic *et al.* [33] attempted to model process variable in the traffic and determined whether the parameter beyond their appropriate ranges using machine learning techniques to detect potential attacks. Sullivan *et al.* [34] added that the appropriate ranges of industrial IoT devices should not only depends on analysis of the traffic, but also need to be revised by the professional and experienced operators. The more suitable learning model for the IDS and IPS system based on the heterogeneous IoT devices still need further study.

*C. Constrained*

*1) Description:* Because of cost and actual physical conditions, many IoT devices like industrial sensor and implantable medical devices have been designed to be lightweight and in small size. Thus, they have much less computing ability and storage resources than traditional computers and mobile phone. In addition, many IoT devices military, industrial, agricultural devices have to work for a long time in environments where charging is not available, so they also have stringent requirements for power consumption. On the other hand, many IoT devices used in vehicle systems, robot control systems and real-time healthcare systems also have to meet the deadline constraints of the real-time processes. We describe the limit resource, power supply and latency of IoT devices as an IoT feature named "constrained" here.

*2) Threats:* Constrained by resource, power supply, and time delay, most IoT devices do not deploy necessary defenses for system and network. For example, lightweight IoT devices do not have the memory management unit (MMU), so memory isolation, address space layout randomization (ASLR) and other memory safety measures cannot be directly deployed on these devices. Moreover, much complicated encryption and authentication algorithms like public cryptography implement on such devices, they occupy too much computing resource and causes a long delay, which affects the normal operation of these devices and reduces performance especially for real-time IoT devices. Consequently, it is easy for attackers to use memory vulnerabilities to compromise these devices. At the same time, due to limit resource many IoT devices even communicate with the server without encryption or use SSL encryption without checking the server's certificate. Attackers could easily intercept communication or launch man-in-the-middle attacks.

*3) Challenges:* How to achieve fine-grain system protections with less system software and hardware resource on lightweight IoT devices is a great challenge for researchers. In addition, such system protections also need to be satisfied the time and power constraints in practical application condition. On the other hand, it is also difficult for researchers to deploy much complex encryption and authentication algorithms with less latency and computing resource on tiny IoT devices.

*4) Solutions & Opportunities:* There are increasing studies focus on designing system security mechanisms for lightweight devices, but most of them still cannot both satisfy the security and application requirements. *ARMor*, [35] a lightweight software fault isolation can be used to sandbox application code running on small embedded processors., but it caused the high-performance overhead for those programs which need checking address many times (e.g. string search). It is not applicable for high real-time demand IoT devices. Koeberl *et al.* [36] presented a set of relatively complete trusted computing functions for lightweight devices such as attestation and trusted execution. However, its implementation has to change the existing hardware architecture of MCU, so it cannot be directly applied to existing IoT devices. Other system defenses like *EPOXY* [37] and *MINION* [38] have been proposed recently better address above challenges, but these protections work base on static analysis of firmware or source code, which will increase the burden on developers.

To protect network security for tiny IoT devices, most cryptology researchers reduce resource consumption by designing new lightweight algorithms [39-41] or optimize the original cryptography algorithms [42]. Nevertheless, it is difficult for lightweight algorithms to achieve the same security level with classical algorithms and new cryptography algorithms may have potential security problems. Some researchers attempt new solutions to address this challenge. For example, Majzoobi team and Hiller team proposed the authentication [43] and key generation algorithm both based on Physical Unclonable Functions (PUF) [44], which use the unique physical structure of the device to identify itself. This method not only saves key resources storage and simplify the algorithm, but also can effectively resist the side channel analysis. Other researchers also tried to use users' unique biological characteristics like gait [45] and usage habits [46] collected by some wearable IoT devices to improve authentication algorithms. It can save resource and authenticate both user and device at same time. However biometric or physical characteristic does not always follow the same pattern. Some unpredictable factors may change them slightly. The stability and the accuracy of these new methods need yet to be further improved.

*D. Myriad*

*1) Description:* Due to the rapidly proliferating IoT devices, the amount of data these devices generated, transited, used will reach be mounting to astronomical figures. We describe the enormous number of IoT devices and the huge amount of IoT data as an IoT feature named "Myriad" here.

*2) Threats:* Last year's Mirai botnet compromised more than

1 million IoT devices, and the attack traffic had exceeded 1Tbps, which previous cyber attacks have never been achieved. Furthermore, more and more new IoT botnet like *IoTroop* [47]. The IoT Botnet was made mostly of unsecured IoT devices rather than computers, and their speed is much faster and would launch large-scale distributed denial of service (DDoS) attacks. Yin *et al.* design honeypot and sandbox system to collect attack samples from IoT devices, and found the most remote network attack use IoT devices launch large scale DDoS attacks [48]. As more IoT applications used in industrial and public infrastructures, the target of IoT botnets would no longer just be the website, but also the important infrastructures, which would bring grave damages to the social security.

*3) Challenges:* Most of IoT devices lack system defense and do not have any safety test software as anti-virus could detect malicious programs. Furthermore, as we discussed before, IoT devices are diversity and very limited in the power supply and computing resource. Thus, how to detect and prevent IoT botnet virus in IoT devices early is great challenge for researchers. At the same time, how to interrupt transmission of huge amount of IoT devices is also a tough problem.

*4) Solutions & Opportunities:* As the increasingly DDoS attack by IoT botnets, many researchers tried to mitigate IoT botnets related cyber risks by using the source code for the Mirai. For instance, JA Jerkins *et al.* [49] designed a tactic that could use the same compromise vector as the Mirai botnet to catalog vulnerable IoT devices, and detect potential poor security practices early. While there still no effective and universal precautions for botnet virus. Zhang and Green [50] first consider the device and environment constraints of IoT network, then design a lightweight algorithm to distinguish malicious requests from legitimate ones in an IoT network, but their assumption was too simple, hackers would not send requests with the same content, but usually simulate users' request with different reasonable content. Moreover, the current DDoS intrusion detection methods only apply in certain scenarios like smart grid [51] or an IoT network based on the single protocol like 6LoWPAN [52].

### E. Unattended

*1) Description:* Smart meters, implantable medical devices (IMDs) and many industrial, agricultural and military sensors in the special physical environment have to perform functions and operate for a long period of time without physical access. As increasing adoption of wireless networking prompts, these devices are evolving into IoT devices. We describe this long-time unattended status of IoT devices as an IoT feature named "unattended" here.

*2) Threats:* In such settings, it is hard to physically connect an external interface to verify the state of these devices. Thus, it is hard to detect when these devices have been remote attacked. In addition, because these devices like IMDs and industrial control devices usually carry out crucial operations, hackers more likely to regard them as prime targets. For example, Stuxnet worm could infect the Programmable Logic Controllers (PLC) used in industrial control systems, which result in considerable physical damage.

*3) Challenges:* As we mentioned above, these "unattended" devices are also made mostly of "constrained" devices. Moreover, they are also usually designed to perform highly specific tasks and interact strongly with the physical environment. Their hardware, system, and process requirements are specific, and it is hard to deploy traditional mobile trusted computing for them [53]. For instance, process memory isolation based on virtual memory is no longer feasible, because many tiny IoT devices are built on hardware that does not provide a memory management unit (MMU). Thus, building trusted execution environment (TEE) to ensure security-critical operations be correctly executed under remote exploits and verifying internal state of a remote unattended tiny IoT device become important tasks in many scenarios.

*4) Solutions & Opportunities: TrustShadow* [75] aims to ensure trusted execution environments for security-critical applications within the context of IoT devices using ARM TrustZone technology. However, such technology is based on the ARM cortex-A processor and does not support tiny IoT devices based on lightweight processor, such as ARM cortex-M. *SMART*, [54] a remote attestation method combing software and hardware to overcome the disadvantages of the only system protection by software or hardware. However, some access control logic of *SMART* like the update of attestation code and interaction between multiple protected modules involve too much delay. Noorman *et al.* [55] built a lightweight trusted execution environment for small embedded, but this method didn't consider how to safely handle the hardware interrupt and memory exception. More effective and widely applicable remote attestation, lightweight trusted execution and safety patch solutions remain open problems.

### F. Intimacy

*1) Description:* As smart meters, wearable devices and even some smart sex toys [56] become more widely used in our lives. These devices not only collect much our biology information including heart rate and blood pressure but also monitor and record our surrounding information and daily activities like the change of indoor temperature and the places you have been. We describe this intimate relationship between users and IoT devices as an IoT feature named "Intimacy" here.

*2) Threats:* The intimate relationships between users and IoT devices will certainly raising more serious and unnoticed privacy concerns. Some researchers [57] show that attackers can infer whether the home is occupied with more than 90 percent accuracy just by analyzing smoke and carbon dioxide sensors data. The power consumption recorded by the smart plug could also be used to analyze your operations on the computers [58]. As cloud-based service will be offered more and more IoT implementations, according to the Gartner Statistics [59]. These sensitive data collected by IoT devices will be shared with service providers. Driven by profit, service providers also keep these data forever and even shared these data with other advertising agency without the user's consent, which can increase the risk of privacy leak. Hackers could obtain the IoT device sensitive data by more sources or acquires illegal benefits by modifying theses data [74].

*3) Challenges:* On the one hand, IoT applications rely on users' personal information to provide service (e.g., auto insurance company collect driving data of each user to offer customized discounts [60]). On the other hand, collecting, transferring and using these sensitive information increases the attack surface of privacy leak. Thus, how to offers an attractive trade-off between sensitive information utility and privacy is a great challenge for the academic community.

*4) Solutions & Opportunities:* Recently, there are increasing studies focusing on the privacy protection of IoT data and anonymous protocols. Many solutions use the data masking and encryption like homomorphic algorithm to protect sensitive information, but these solutions reduce the availability of original data and increase the time delay. Effective privacy protection method should protect users' privacy, remain high availability of original data and guarantee real-time at the same time. Another major problem of current privacy protection method is narrow application scope and incomprehensive protection. Most solutions only applied to a certain application scenarios, (e.g., smart grid [61], smart medical [62] or car networking [63]), or to the specific process of data lifecycle (e.g., data collection [64], privacy data sharing with the cloud service [65]). More complete and general protection needs more in-depth research, including data collection, transmission, use, storage, and sharing.

Conversely, due biological characteristics are different from person to person, the intimate relationships between users and IoT devices could also be contributed to cryptography. As we discussed above researchers could use these biological signals collected by devices to generate a unique encryption key for users or to provide authentication [66].

*G. Mobile*

*1) Description:* Many IoT devices as wearable devices and smart cars are used in the mobile environment. These mobile devices often need to hop from one network environment to another environment and have to communicate with many unknown new devices. For example, use drive smart car from one district to another, the car will automatically collect road information for highway foundational facilities in the new district. This scenario will be more common in the future of social IoT. We describe the movement of IoT devices as an IoT feature named "mobile" here.

*2) Threats:* Because mobile IoT devices are more likely to join more networks, hackers tend to inject the malicious code into mobile IoT devices to accelerate the spread of malicious code. At the same time, because mobile devices need to communicate with more devices, the attack surface of mobile themselves will be border. The coming crisis tend to be worse in social IoT devices. In future, the social IoT devices would carry more sensitive information and automatically follow the user's joining from one social network into another.

*3) Challenges:* In response to the threat, the main security challenge should be addressed is cross-domain identification and trust. For example, when a mobile device hops from one domain to another and how the new domain to verify this device and what kind of permissions should give to it. When data carried with mobile devices passed from one network or protocol to another, it also involves key negotiation, data confidentiality, integrity protection and other important security issues.

*4) Solutions & Opportunities:* Chen *et al.* [67] try to decrease the probability of being attacked by dynamically changing the configuration of devices according to the trust condition of other devices in different networks. This method would not address the root of the problem. There are few suitable access control policies for the mobile devices have been proposed. More thorough studies should be done to solve these problems in this area.

*H. Ubiquitous*

*1) Description:* The IoT devices have pervaded every aspect of our lives. We will not just use them, but also rely on them and even be more dependent than the smartphone. IoT will become an indispensable part of people's daily lives like air and water. We describe the phenomena that IoT devices will be everywhere in our future lives as an IoT feature named "Ubiquitous". In this section, we do not focus on this feature effect on security in the technology as above. We will discuss the lack of security and privacy awareness of the "ubiquitous" IoT devices and its resulting threat. We will also give some suggestions should adopt towards the "ubiquitous" IoT devices. In addition, we will discuss above issues from the following four distinct social roles: ordinary consumers, manufacturers, professional operators, and security researchers.

*2) Threats & Suggestions*

   *a) Consumers:* As the IoT device is taking off in emerging markets, the number of devices will surpass the number of humans. According to the statistics from Govtech [68], everyone will own an average of six to eight IoT devices by 2020. That is just the number of the devices everyone owns, and the number of the actual devices everyone use will be more. However, most people still lack the management awareness and privacy protection awareness. As IoT devices more intelligent and closer to our lives, they could automatically complete many assignments without any manual intervention and even any reminders. Thus, many users do not realize their devices have been compromised until attackers lead to more obvious and serious consequences. People always ignore the safety and reliability of IoT devices when buying and using IoT products. Therefore, malware like Mirai virus can just use default username and password to perform remote control so many IoT devices. In 2014, WeLiveSecurity team highlighted the discovery of 73,000 security cameras with default passwords [69]. Consumers should change their consciousness from a user to an administrator and pay attention to IoT security as the same way to food safety. Only in this way could we fundamentally avoid "human" becoming the weakest link in the IoT security.

   *b) Manufacturers:* The IoT device manufacturers also do not attach enough importance to the security of IoT products. A large proportion of manufacturers consider security measures will add additional cost without any profits. Thus, company keeps producing and deploying new IoT devices with insecure-by-default configuration. These devices not only have many known vulnerabilities, but also have the potential flaw in their design. For example, the In-Vehicle infotainment systems or vehicle navigation systems in many smart cars directly connect to CAN-Bus. Attackers could compromise these systems, and then use the CAN-Bus to control the car [70], as shown in Fig. 4.

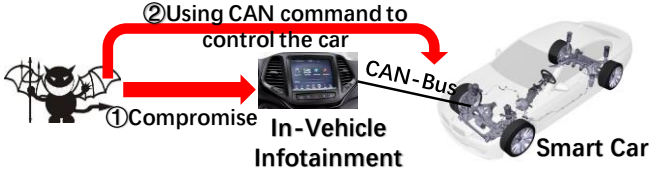

Fig. 4. Attack Example of Insecure Configuration.

On the other hand, enterprises usually do not supply any security service for customers. For example, manufacturers always only write simple instructions in their manual without any security suggestions and notices. Customer usually could not know what sensitive information the devices will collect, and how to more safely use them. Manufacturers also do not take the initiative to help customers install patches or update firmware against new malware threats and even do not send any security warnings. Therefore, IoT devices vulnerabilities have longer exploited period and broader impact than traditional computer vulnerabilities. It is the urgent needs of setting the detailed security standards for IoT products. IoT manufactures also should work tightly with the supervisory agencies, as DHS and FSA.

*c) Operators:* With the IoT devices are widely used in industry, agriculture and even military fields, the security awareness of profession operators also needs to be raised. Most operators consider [71] attackers may do not know how to use these specialist devices, let alone attack them. Thus, when these devices have abnormal behaviors, most operators' first response is the malfunction of the equipment or the failure of their own operations. However, attacking a well-targeted device is much easier than using all devices correctly, thus operators should increase the sensitivity of abnormal behaviors and must be skilled in using security tools like IDS and IPS.

*d) Researchers:* As IoT devices are applied to more scenarios, there will be more types and functions of devices with different resources and architectures, as we mentioned above. Researchers should no longer only focus on theory study, and need more cooperation with consumers, manufacturers and professional operators. Then researchers could have more comprehensive insight into the actual usage of IoT devices in the real conditions and design more practical safety precautions with fewer resource demand and lower extra cost.

*I. Summary*

The features we discussed above are not independent but interact with each other. For instance, the resource of most unattended devices is constrained. When designing security solutions for these devices, researchers need to take the effect of both features into consideration. In addition, other IoT features that have less impact on security and privacy are out of the scope. Also, some IoT features such as extensibility and integration may bring certain security and privacy issues, but most of these issues have much overlap with the features we have discussed above. We finally summarized the main threats, challenges, and opportunities of each feature in Table I.

TABLE I
THREATS, CHALLENGES, AND OPPORTUNITIES OF EACH IoT FEATURES

| Feature | Threat | Challenge | Opportunity |
| --- | --- | --- | --- |
| *Inter-dependence* | Bypassing static defenses, Overprivilege | Access control and privilege management | Context-based permission |
| *Diversity* | Insecure protocols | Fragmented | Dynamic analysis simulation platform, IDS |
| *Constrained* | Insecure systems | Lightweight defenses and protocols | Combining biological and physical characteristics |
| *Myriad* | IoT botnet, DDoS | Intrusion detection and prevention | IDS |
| *Unattended* | Remote attack | Remote verification | Remote attestation, Lightweight trusted execution |
| *Intimacy* | Privacy leak | Privacy protection | Homomorphic encryption, Anonymous protocols |
| *Mobile* | Malware propagation | Cross-domain identification and trust | Dynamic configuration |
| *Ubiquitous* | Insecure configuration | \ | Safety consciousness |

## III. IoT SECURITY RESEARCH ANALYSIS

In order to grasp the latest trend of development of IoT security research and better understand how above IoT features affect existing security research, we studied nearly 200 research papers related to IoT security from top journals and conferences in recent five years. We will illustrate the development of IoT security research base on these research and reveal the reasons behind it. We also give some suggestions to researchers based on the analysis and help them to keep up with the latest IoT security research status and research priorities for further study.

*A. Research Collection and Label*

To help with understanding the statistical analysis and classification of IoT research papers in the remainder of this section. We first explain how we searched and filtered existing research papers either in or out of our study scope, and introduce how we labeled each paper in this section.

Firstly, we collected the research paper from leading journals and conferences in computer security (concrete catalog see the GitHub link in Appendix). Then we determined whether the research is related IoT security by following procedure. Firstly, we chose some words directly related to IoT as IoT keywords including all kinds IoT devices, protocols and application scenarios (e.g., smart watch, smart home, WSN). Then if the title of paper contains these IoT keywords or its abbreviation, we added it to our study list. Otherwise, we checked whether the abstract of this paper includes the word "privacy" or "security", and IoT keywords at the same time. Finally, there were nearly 200 research papers selected for further study (all tags of these papers see the GitHub link in Appendix).

In addition, in order to classify these papers according to SOA IoT layers (e.g., sensing, transfer, service, and interface) [72] or application scenarios for the further statistical analysis and to find what problem research care most at present stage, we labeled three tags (layers, application scenarios and threat) with every paper. It is easy to determine which layer and application the paper belong to base on its topic. Although the solutions of these papers are different from each other, what problems of some solutions try to solve are close. Thus, we label which "threat" tag of each paper base on these common problems. To better generalize common problems, we describe these problems mainly base on OWASP IoT Top Ten [73] security issues (e.g., privacy concern and vulnerable cloud service).

### B. Statistical Analysis

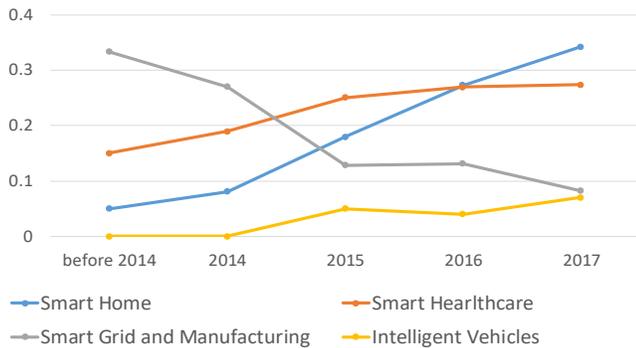

Fig. 5. The proportion of the Number of Papers in Different Application Scenarios per Year

The Fig. 5 illustrates the change of the proportion of the number of papers in some application scenarios in recent years. We can find the IoT security research hotspot always follows the development of IoT applications. For example, in the early 2010s, smart grid and smart manufacturing got more popularization and application, thus the security research in these fields are more than others. However, with the rapid development of smart home and healthcare technology over the last three years, security researchers turned more attention to these field, at the same time, the research interests in the smart grid and smart manufacturing was on the decline.

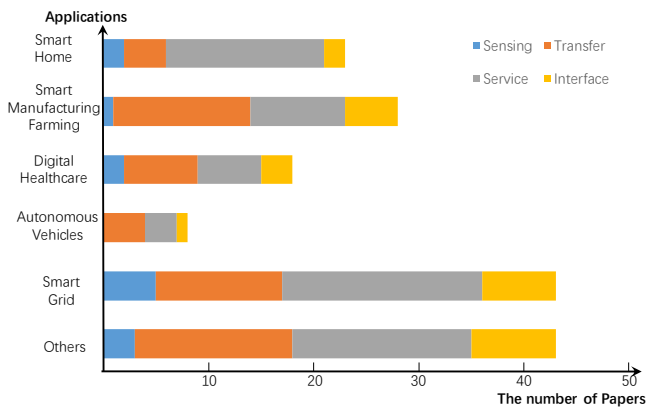

Fig. 6. The Number of Papers of Each Layer in Different IoT Application Scenarios

> SUGGESTION: Security researchers should pay attention to the new IoT applications, to prevent the potential threats before they emerge.

The Fig. 6 shows the number of research papers in each layer of every IoT application scenario. As can be seen from the figure security studies distribution of different layers varied from one application scenario to another. For instance, there are more research of transfer layers in smart manufacturing than in application layer, but it is opposite in smart home. That is because in industrial and agriculture environment, all sensors depend on wireless sensor network (WSN) to communicate with each other and remote control system. The security problems in WSN will be more dangerous to others. By contrast, smart home devices are controlled by mobile applications or web applications. Thus, more researchers drew more attention to application security in smart home, and transfer security in smart manufacturing. Fig. 7 also reinforces this view.

> SUGGESTION: IoT devices in different IoT application scenarios have different working models. Researchers should understand the differences between different application scenarios to grasp their main security problems.

We counted the number of research papers of each "threat" tag in every application scenario, as shown in Fig. 7. Most of the research efforts have been focused on migrating privacy disclosure and insecure network or protocol problems. That is just due to the "intimacy", "myriad", and "diversity" features which we have discussed above. More sensitive information has been collected, transferred and used by IoT devices especially smart home and healthcare devices, which must involve more privacy issues. Due to a large number of IoT devices, attackers more easily to carry out cyber attacks. Most new kinds of devices and protocols also have many vulnerabilities which catching more efforts to solve these problems. The main reason for casing insufficient security configures and vulnerable cloud and web service is the lack of awareness as we mentioned above. In addition, although research on IoT system and IoT mobile application are less in the past years, more attackers will find and use the potential system and application vulnerabilities caused by the "constrained" and "interdependence" IoT features. More studies should work in these fields.

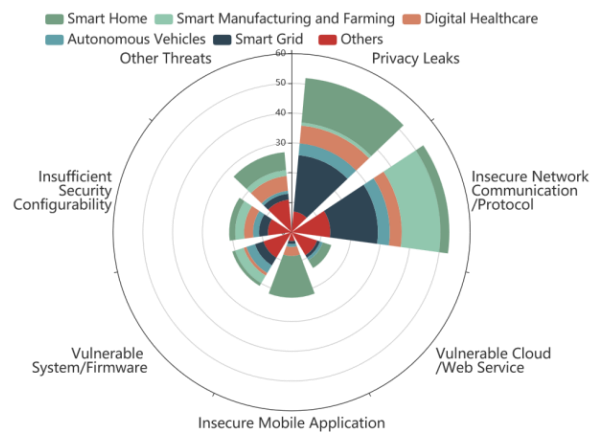

Fig. 7. The Number of Papers of Different Threat Tags in Different Application Scenarios

SUGGESTION: Researchers need to investigate further to discover the root causes and new IoT features behind new security threats, and design more generic and practical protective measures.

## IV. CONCLUSION

In this paper, we analyzed and discussed the security and privacy issues base on IoT features. We first presented what the threats and research challenges born from these features. Then we also studied existing solutions for these challenges and pointed out what new security technology required further. Finally, we illustrated the development trend of recent IoT security research, the reason for it, and how IoT features reflect on the existing research. Only by deeply analyzing these new features behind the Internet of things, we can get a better idea about the future research hotspots and development of the IoT security.

APPENDIX

We publish all research and survey papers that we collected and studied on the GitHub as shown below. We will continue to update our research papers.

*https://github.com/chaojixx/IoT-security-papers*

REFERENCES

[1] The Statistics Portal. (2017). *Internet of Things (IoT) connected devices installed base worldwide from 2015 to 2025 (in billions).* [Online]. Available: https://www.statista.com/statistics/471264/iot-number-of-connected-devices-worldwide/
[2] IDC. (2016). *Internet of Things Market Statistics.* [Online]. Available: http://www.ironpaper.com/webintel/articles/internet-of-things-market-statistics/
[3] Bigthink Edge. (2016). *Hacking the Human Heart* [Online]. Available: http://bigthink.com/future-crimes/hacking-the-human-heart
[4] Envista Forensics. (2015). *The Most Hackable Cars on the Road.* [Online]. Available: http://www.envistaforensics.com/news/the-most-hackable-cars-on-the-road-1
[5] Wikipedia. *2016 Dyn cyberattack.* [Online]. Available: https://en.wikipedia.org/w/index.php?title=2016_Dyn_cyberattack&oldid=763071700
[6] Langner, Ralph. "Stuxnet: Dissecting a Cyberwarfare Weapon." IEEE Security & Privacy 9.3(2011):49-51.
[7] Richard Patterson. (2017). *How safe is your data with the IoT and smart devices.* [Online]. Available: https://www.comparitech.com/blog/information-security/iot-data-safety-privacy-hackers/
[8] GeekPwn. (2017). *IoT devices have a large number of low-level loopholes.* [Online]. Available: http://www.sohu.com/a/129188339_198147
[9] Li, Shancang, T. Tryfonas, and H. Li. "The Internet of Things: a security point of view." Internet Research 26.2(2016):337-359.
[10] Lin, Jie, *et al.* "A Survey on Internet of Things: Architecture, Enabling Technologies, Security and Privacy, and Applications." *IEEE Internet of Things Journal.*, vol. 99, p1 2017.
[11] Fu, Kevin, *et al.* (2017). S*afety, Security, and Privacy Threats Posed by Accelerating Trends in the Internet of Things. Technical Report. Computing Community Consortium.* [Online]. Available: http://cra.org/ccc/wp-content/uploads/sites/2/2017/02/Safety-Security-and-Privacy-Threats-in-IoT. pdf.
[12] R. Roman, J. Zhou, and J. Lopez, "On the features and challenges of security and privacy in distributed Internet of Things," *Comput. Netw.*, vol. 57, no. 10, pp. 2266–2279, 2013.
[13] Sicari, S., *et al.* "Security, privacy and trust in Internet of Things: The road ahead." *Computer Networks the International Journal of Computer & Telecommunications Networking* 76.C (2015):146-164.
[14] Yang, Yuchen, *et al.* "A Survey on Security and Privacy Issues in Internet-of-Things." IEEE Internet of Things Journal 4.5(2017):1250-1258.
[15] W. Trappe, R. Howard, and R. S. Moore, "Low-energy security: Limits and opportunities in the Internet of Things," *IEEE Security Privacy*, vol. 13, no. 1, pp. 14–21, Jan./Feb. 2015
[16] Linden Tibbets and Jesse Tane. (2012). *IFTTT.* [Online]. Available: https://platform.ifttt.com/
[17] Samsung. (2014). *SmartThings.* [Online]. Available: https://www.smartthings.com/
[18] Apple. (2014). *HomeKit.* [Online]. Available: https://developer.apple.com/homekit/
[19] Amazon. (2012). *Alexa.* [Online]. Available: https://developer.amazon.com/alexa
[20] Fernandes, Earlence, J. Jung, and A. Prakash. "Security Analysis of Emerging Smart Home Applications." *Security and Privacy IEEE*, 2016, pp. 636-654.
[21] Yu, Tianlong, *et al.* "Handling a trillion (unfixable) flaws on a billion devices: Rethinking network security for the Internet-of-Things." *ACM Workshop on Hot Topics in Networks*, 2015, pp. 5.
[22] Jia, Yunhan Jack, *et al.* "ContexIoT: Towards Providing Contextual Integrity to Appified IoT Platforms." *Network and Distributed System Security Symposium* 2017, pp. 1-15.
[23] JD. (2015). *Alpha-IoT.* [Online]. Available: http://devsmart.jd.com/
[24] Ali. (2015). *Alibaba Smart Living.* [Online]. Available: https://www.aliplus.com/
[25] Ali. (2015). *Internet of things security report.* [Online]. Available: https://jaq.alibaba.com/community/art/show?articleid=195
[26] Network Working Group Internet-Draft. (2017). *Secure IoT Bootstrapping: A Survey.* [Online]. Available: https://tools.ietf.org/html/draft-sarikaya-t2trg-sbootstrapping-03
[27] Liu, Hui, *et al.* "Smart Solution, Poor Protection: An Empirical Study of Security and Privacy Issues in Developing and Deploying Smart Home Devices." *IoT Security & Privacy Workshop 2017*, pp. 13-18.
[28] Yang Yu. *BadTunnel: NetBIOS Name Service spoofing over the Internet* [Online]. Available: https://www.blackhat.com/docs/us-16/materials/us-16-Yu-BadTunnel-How-Do-I-Get-Big-Brother-Power-wp.pdf
[29] Rubio-Hernan, Jose, J. Rodolfo-Mejias, and J. Garcia-Alfaro. "Security of Cyber-Physical Systems." *Conference on Security of Industrial-Control- and Cyber-Physical Systems Springer*, Cham, 2016, pp. 3-18.
[30] Davidson, Drew, *et al.* "FIE on Firmware: Finding Vulnerabilities in Embedded Systems Using Symbolic Execution." *USENIX Security Symposium*. 2013, pp. 463-478.
[31] Zaddach, Jonas, *et al.* "AVATAR: A Framework to Support Dynamic Security Analysis of Embedded Systems' Firmwares." NDSS. 2014.
[32] Chen, Daming D., *et al.* "Towards Automated Dynamic Analysis for Linux-based Embedded Firmware." *Network and Distributed System Security Symposium*. 2016.
[33] Hadžiosmanović, Dina, *et al*. "Through the eye of the PLC." The, Computer Security Applications Conference 2014, pp. 126-135.
[34] Sullivan, Daniel T., and Edward J. Colbert. Network Analysis of Reconnaissance and Intrusion of an Industrial Control System. No. ARL-TR-7775. Computational and Information Sciences Directorate, US Army Research Laboratory Adelphi United States, 2016.
[35] Zhao, Lu, *et al.* "ARMor: fully verified software fault isolation." *Proceedings of the International Conference on Embedded Software IEEE*, 2011:289-298.
[36] Schulz, Patrick Koeberl Steffen, Ahmad-Reza Sadeghi, and Vijay Varadharajan. "Trustlite: A security architecture for tiny embedded devices." *EuroSys*. ACM, 2014, pp:1-14.
[37] Clements, Abraham A., *et al.* "Protecting Bare-Metal Embedded Systems with Privilege Overlays." *Security and Privacy IEEE*, 2017.
[38] Chung, Taegyu.,*et al.* "Securing Real-Time Microcontroller Systems through Customized Memory View Switching." *Network and Distributed System Security Symposium*, 2018.
[39] Guo, Fuchun, *et al.* "CP-ABE With Constant-Size Keys for Lightweight Devices." *IEEE Transactions on Information Forensics & Security 9.5*. 2014, pp. 763-771
[40] Fan, Hongfei, *et al.* "An ultra-lightweight white-box encryption scheme for securing resource-constrained IoT devices." Conference on Computer Security Applications ACM, 2016, pp.16-29.
[41] Buchmann, Johannes, *et al.* "High-performance and lightweight lattice-based public-key encryption." *Proceedings of the 2nd ACM*

*International Workshop on IoT Privacy, Trust, and Security*. ACM, 2016, pp. 2-9.

[42] Rauter, Tobias, N. Kajtazovic, and C. Kreiner. "Privilege-Based Remote Attestation: Towards Integrity Assurance for Lightweight Clients." *ACM Workshop on IoT Privacy, Trust, and Security* .ACM, 2015, pp. 3-9.

[43] Majzoobi, Mehrdad, *et al.* "Slender PUF Protocol: A Lightweight, Robust, and Secure Authentication by Substring Matching." *Security and Privacy Workshops IEEE*, 2012. pp. 33-44.

[44] Hiller, Matthias, G. Sigl, and M. Bossert. "Online Reliability Testing for PUF Key Derivation." *International Workshop on Trustworthy Embedded Devices*. ACM, 2016, pp.:15-22.

[45] Xu, Weitao, *et al.* "KEH-Gait: Towards a Mobile Healthcare User Authentication System by Kinetic Energy Harvesting." *The Network and Distributed System Security Symposium*. 2017.

[46] Scheel, Ryan A., and A. Tyagi. "Characterizing Composite User-Device Touchscreen Physical Unclonable Functions (PUFs) for Mobile Device Authentication." *International Workshop on Trustworthy Embedded Devices*. ACM, 2015, pp. 3-13.

[47] Checkpoint Research. (2017). *IoTroop Botnet: The Full Investigation.* [Online]. Available: https://research.checkpoint.com/iotroop-botnet-full-investigation/

[48] Yin, Minn Pa Pa, *et al.* "IoTPOT: analysing the rise of IoT compromises." *Usenix Conference on Offensive Technologies*. USENIX Association, 2015, pp. 9-9.

[49] Kolias, Constantinos, *et al.* "DDoS in the IoT: Mirai and Other Botnets." *Computer*. vol. 50, no. 7, pp. 80-84, 2017.

[50] Zhang, Congyingzi, and R. Green. "Communication security in internet of thing: preventive measure and avoid DDoS attack over IoT network." *Symposium on Communications & NETWORKING Society for Computer Simulation International*, 2015, pp. 8-15.

[51] Lu, Zhuo, W. Wang, and C. Wang. "Camouflage Traffic: Minimizing Message Delay for Smart Grid Applications under Jamming." *Dependable & Secure Computing IEEE Transactions*, vol. 12 no.1, pp. 31-44, 2015.

[52] Kasinathan, Prabhakaran, *et al.* "DEMO: An IDS framework for internet of things empowered by 6LoWPAN." *ACM Sigsac Conference on Computer & Communications Security*. ACM, 2013, pp. 1337-1340.

[53] Rubio-Hernan, Jose, J. Rodolfo-Mejias, and J. Garcia-Alfaro. "Security of Cyber-Physical Systems." *Conference on Security of Industrial-Control- and Cyber-Physical Systems Springer*, Cham, 2016, pp. 3-18.

[54] K. E. Defrawy, A. Francillon, D. Perito, and G. Tsudik, ''SMART: Secure and minimal architecture for (establishing a dynamic) root of trust,'' *Network. & Distribution. System. Security Symp.*, 2012.

[55] J. Noorman *et al.*, ''Sancus: Low-cost trustworthy extensible networked devices with a zero-software trusted computing base,'' *22nd USENIX Conf. Security*, 2013, pp. 479–494.

[56] Elizabeth Armstrong Moore. USA Today. 2016. *Woman sues sex-toy maker for invading privacy.* [Online]. Available: http://www.usatoday.com/story/news/2016/09/15/womansues-sex-toy-maker-invading-privacy/90400592/. (2016)

[57] Copos, Bogdan, *et al.* "Is Anybody Home? Inferring Activity From Smart Home Network Traffic." *Security and Privacy Workshops IEEE*, 2016, pp. 245-251.

[58] Nati, Michele, *et al.* "Mind The Plug! Laptop-User Recognition Through Power Consumption." *ACM International Workshop on IoT Privacy, Trust, and Security*. ACM, 2016, pp. 37-44.

[59] Volansys. (2016) *Connecting Devices to Cloud IoT Platform-as-a-Service: Challenges and Solution.* [Online]. Available: https://volansys.com/connecting-devices-cloud-iot-platform-service-challenges-solution/

[60] MarketsandMarkets. (2015, Dec). *Insurance Telematics Market Worth 2.21 Billion USD by 2020.* [Online]. Available: http://www.prnewswire.com/news-releases/insurance-telematics-market-worth-221-billion-usd-by-2020-561817961.html

[61] Yang, Weining, *et al.* "Minimizing private data disclosures in the smart grid." *ACM Conference on Computer and Communications Security*. ACM, 2012, pp. 415-427.

[62] Chan, Ellick M., P. E. Lam, and J. C. Mitchell. "Understanding the challenges with medical data segmentation for privacy." *Usenix Conference on Safety, Security, Privacy and Interoperability of Health Information Technologies*. USENIX Association, 2013, pp. 2-2.

[63] Guo, Longhua, *et al.* "A Secure Mechanism for Big Data Collection in Large Scale Internet of Vehicles." *IEEE Internet of Things Journal*. vol. 99 pp.1, 2017

[64] Barthe, Gilles, *et al.* "Verified Computational Differential Privacy with Applications to Smart Metering." *Computer Security Foundations Symposium IEEE*, 2013, pp. 287-301.

[65] Li, Fengjun, F. Li, and F. Li. "A multi-cloud based privacy-preserving data publishing scheme for the internet of things." *Conference on Computer Security Applications ACM*, 2016, pp. 30-39.

[66] Chang, Sang Yoon, *et al.* "Body area network security: robust key establishment using human body channel." *Usenix Conference on Health Security and Privacy*. USENIX Association, 2012, pp.5.

[67] Chen, Ing Ray, F. Bao, and J. Guo. Trust-based Service Management for Social Internet of Things Systems. IEEE Computer Society Press, 2016.

[68] Govtech. (2015). *FutureStructure: the new framework for communities (Infographic).* [Online]. Available: http://www.govtech.com/dc/articles/FutureStructure-The-NewFramework-for-Communities.html

[69] WeLiveSecurity.(2016, October). *10 things to know about the October 21 IoT DDoS attacks* [Online]. Available: https://www.welivesecurity.com/2016/10/24/10-things-know-october-21-iot-ddos-attacks/

[70] Miller, Charlie, and Chris Valasek. "Remote exploitation of an unaltered passenger vehicle." *Black Hat*, USA, 2015.

[71] Wright, Alex. "Mapping the internet of things." *Communications of the ACM*, vol. 60, no.1, pp. 16-18, 2016.

[72] Bi, Z., Xu, L., and Wang, C. (2014), "Internet of Things for Enterprise Systems of Modern Manufacturing," *IEEE Transactions on Industrial Informatics*, Vol. 10, No. 2, pp. 1537 - 1546 2014

[73] OWASP. (2014). *OWASP Internet of Things Top Ten*. [Online]. Available: https://www.owasp.org/images/7/71/Internet_of_Things_Top_Ten_2014-OWASP.pdf.

[74] Le Guan, Jun Xu, Shuai Wang, Xinyu Xing, Lin Lin, Heqing Huang, Peng Liu and Wenke Lee, "From Physical to Cyber: Escalating Protection for Personalized Auto Insurance," in Proceedings of the 14th ACM Conference on Embedded Network Sensor Systems, SenSys '16, pp. 42-55, 2016.

[75] Guan, Le, et al. "TrustShadow: Secure execution of unmodified applications with ARM trustzone." Proceedings of the 15th Annual International Conference on Mobile Systems, Applications, and Services. ACM, 2017.

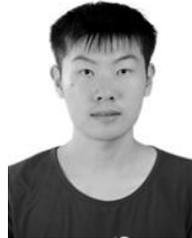

**Wei Zhou** received the BS degree in information security in 2012 from Xidian University, Xian, Shaanxi, China. He is currently pursuing the Ph.D. degree at the National Computer Network Intrusion Protection Center at University of Chinese Academy of Sciences, Beijing, China, under the supervision of Dr. Zhang.

His research interests include the fields of embedded system security, trust computing, network security.

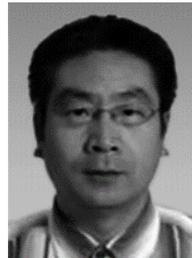

**Yuqing Zhang** received his Ph.D. degree in Cryptography from Xidian University, China. Dr. Zhang is a Professor and the Director of the National Computer Network Intrusion Protection Center at University of Chinese Academy of Sciences.

His research interests include network and system security, and applied cryptography. He has published more than 100 research papers in international journals and conferences, such as ACM CCS, IEEE TPDS, and IEEE TDSC. His research has been sponsored by NSFC, Huawei, Qihu360, and Google. He has served as program chair for more than 5 international workshops (e.g.,

SMCN-2017), and PC member for more than 10 international conferences in networking and security, such as IEEE Globecom 16/17, IEEE CNS 17, and IFIP DBSec 17.

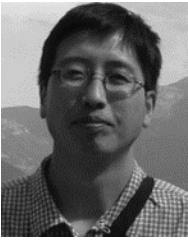 **Peng Liu** received the BS and MS degrees from the University of Science and Technology of China and the Ph.D. degree from George Mason University, in 1999.

He is a professor of information sciences and technology, founding director of the Center for Cyber-Security, Information Privacy, and Trust, and founding director of the Cyber Security Lab, Penn State University. His research interests include all areas of computer and network security.

He has published a monograph and more than 260 refereed technical papers. His research has been sponsored by US National Science Foundation, ARO, AFOSR, DARPA, DHS, DOE, AFRL, NSA, TTC, CISCO, and HP. He has served on more than 100 program committees and reviewed papers for numerous journals. He received the DOE Early Career Principle Investigator Award. He has co-led the effort to make Penn State an NSA-certified National Center of Excellence in Information Assurance Education and Research. He has advised or co-advised more than 30 Ph.D. dissertations to completion. He is a member of the IEEE